\documentclass[prc,aps,twocolumn]{revtex4}
\usepackage{graphicx} 
\usepackage{amsmath}
\usepackage{slashed}
\usepackage{ulem}
\usepackage{hyperref}
\usepackage{color}
\usepackage{url}

\begin{document}

\title{A Novel Feature of Valence Quark Distributions in Hadrons}

\author{Christopher Leon, Misak M. Sargsian and Frank Vera}

\affiliation{Department of Physics, Florida International University, Miami, FL 33199 USA}

\date{\today}

 \begin{abstract}
Examining the  evolution of the maximum of valence quark distribution weighted by Bjorken x,   $h(x,t)\equiv xq_V(x,t)$, 
we observe that  $h(x,t)$ at the peak  should become a one parameter function;   
$h(x_p,t)=\Phi(x_p(t))$,  where $x_p$ is the position of the peak and  $t= \log{Q^2}$. 
This observation is used to derive a  new model independent relation which
connects  the  partial  derivative of the valence parton distribution functions (PDFs) in $x_p$  to the 
QCD evolution equation through the $x_p$-derivative of  the logarithm of  the function $\Phi(x_p(t))$.
 A numerical analysis of this relation using empirical
 PDFs results in a observation of the 
exponential form of the $\Phi(x_p(t)) = h(x_p,t) = Ce^{D x_p(t)}$ for leading to next-to-next leading order 
 approximations of PDFs for  the all  $Q^2$ range covering four orders  in magnitude.  
The exponent, $D$,  of the observed ``height-position" correlation function  converges with the increase of 
the  order of approximation.  This result holds for all PDF sets considered.
A  similar relation is observed also for  pion valence quark distribution, indicating  that the obtained relation may be  
universal for any non-singlet partonic distribution.
The observed ``height - position"  correlation is used also to indicate that no finite number exchanges 
can describe  the analytic behavior of the valence quark distribution at the position of the peak
at fixed $Q^2$.
\end{abstract}
\maketitle


\section{ Introduction} 

Valence quarks play a unique role in the QCD  dynamics of  hadrons.
They define the baryonic number of the nucleons and represent ``effective'' fermions  with  complex 
interactions among themselves and with the hadronic interior.  One important property of valence quarks is that 
their number is conserved and  hadrons can be considered as systems with 
fixed number of effective (valence) fermions.
The continuing progress in  experimental extraction of valence quark distributions in a wide range of $x$
and the  emerging new possibilities for probing them in semi-inclusive and exclusive deep-inelastic processes  
create a new motivation  for theoretical  modeling of their dynamics.  This modeling is important since in the 
case of success one gains a new level of understanding of the QCD dynamics in the hadrons. 
Even if lattice calculations can reproduce the major characteristics of valence quark distributions they don't necessarily 
result in a qualitative understanding of the  underlying processes.
In this respect, observation of new properties and  relations in valence quark  distributions  is significant  since it allows one 
to constrain models aimed at describing the dynamics of  QCD interaction.

The possibility of considering nucleons as a system of  ``effective fermions"  (i.e. valence quarks) whose number is conserved, 
opens a new venue in exploration of the dynamics of the valence quarks from the point of  view of universal properties 
of two component (spin or isospin) fermi systems with conserved number of constituents.
This  approach is analogous to the recent study of the ultra cold two component fermi atomic systems and 
atomic nuclei for which, despite 20 orders of magnitude difference in the density, a similar analytic 
form for the high momentum tail of the momentum distribution was found\cite{Hen:2014lia} based on the universal properties of 
the fermi system. We follow a similar logic in  studying QCD structure of hadrons and 
our focus  in the present work is on one of the  most distinguishable  characteristics of valence quarks which  is, 
their distribution, $q_V(x,Q^2)$ weighted by momentum fraction, $x$, exhibits  a clear peak\footnote{
Note that  furthermore we will  refer to $xq_V(x,Q^2)$ as a quark structure function, which is a renormalization/factorization scheme dependent quantity that evolves with scale and should not be confused with the structure function of hadrons, which are observables.}.
This peak is  a hallmark for the bound system of conserved number fermions\footnote{No such peak exists for 
sea quark distribution.} and is characterized by its height, $h(x_p)$,
and the position, $x_p$, both of which 
evolve with the resolution scale, $Q^2$.
  \begin{figure}[th]
 \vspace{-0.4cm}
\centerline{ \includegraphics[width=10cm,height=3.6cm]{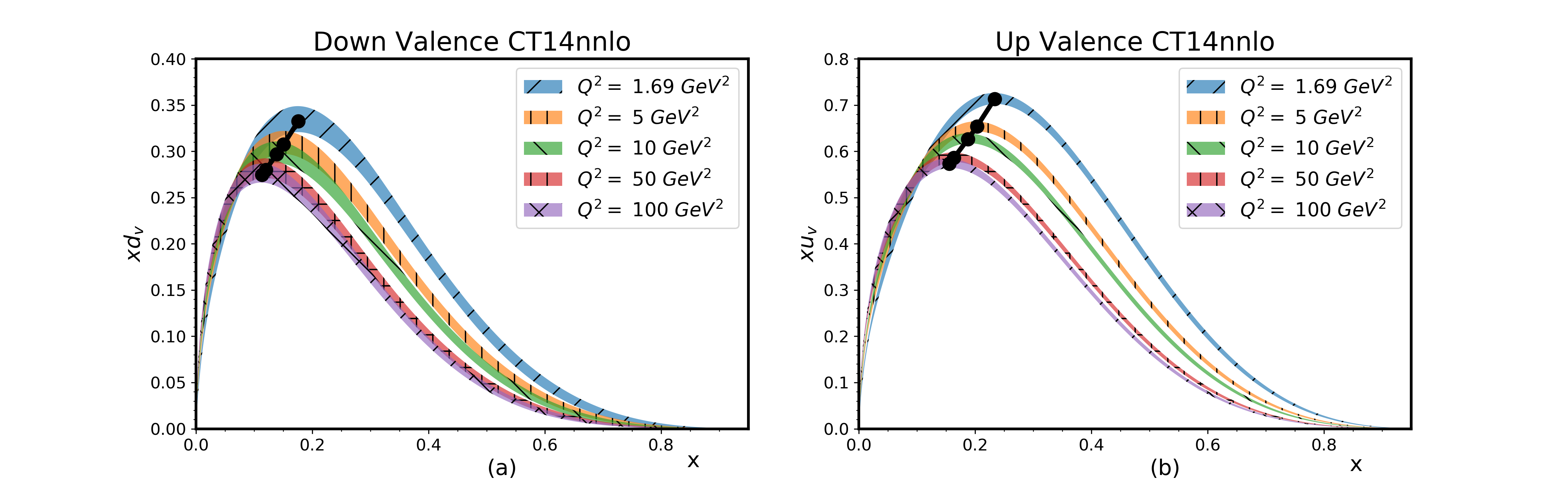}}
 \vspace{-0.6cm}
 \caption{(Color online.)  The $x$ dependence of $xq_V(x,t)$ distributions at different values of $Q^2$ for down and up 
 valence quarks in proton. 
 Peaks at different $Q^2$ are connected by dashed line to   visualize the correlation between the position and the height of the peak. The shaded area is the Hessian error at 68\% confidence level.}
\label{peaksandposes}
\end{figure}
 
Since this  peak is a  product of the  magnitude of $x$ and the strength  of the valence quark distribution one expects that 
its position and the height to be $Q^2$ dependent  due to the QCD evolution of the  valence quark distribution,  
$q_V(x,Q^2)$, whose strength shifts towards smaller $x$  with increasing $Q^2$.  As a result one expects that 
both  the position of the peak, $x_p$ and its height, $h(x_p,t) =  x_p(t) q_V(x_p,t)$ to  be a function of  
$Q^2$ (hereafter we use the variable, $t = \log{Q^2}$).

\section{``Height-Position" Correlation of  the Peak of  the Valence Quark Structure Functions} 

As Fig.\ref{peaksandposes} shows the height of the peak and its position  for valence PDFs in the nucleon diminishes with an increase of $Q^2$ 
as one expects from the QCD evolution that moves the strength of PDFs towards small $x$.
If now we assume that both the {\it height}  and  the {\it position} of the peak of the valence 
quark structure function evolve due to the evolution of  the strong coupling, then these dimensionless 
quantities  can be expressed as:  
\begin{equation}
x_p\left({Q^2\over \mu^2}, \alpha_s(\mu)\right)    =     x_p(1, \alpha_s(Q))  =    \sum\limits_{n=0}^\infty x_n \alpha^n_s(t)     
\label{xp(al)}
\end{equation}
\begin{equation}
h\left({Q^2\over \mu^2}, \alpha_s(\mu)\right)   =   h\left(1, \alpha_s(Q)\right) = 
\sum\limits_{n=0}^\infty h_n \alpha_s^n(t),
\label{hp(al)}
\end{equation}
where $x_n$ and $h_n$ are constants and $\alpha_s(t)$ is the strong interaction coupling constant evaluated at $t=\log{Q^2}$.

Eq.(\ref{xp(al)})  is a single variable  function of $\alpha_s$, continuously differentiable with non zero
derivate.
Thus it is  in general an invertible function.  This fact allows us to combine \  Eqs.~(\ref{xp(al)}) and 
(\ref{hp(al)})  representing  the height, $h(x_p,t)$   as one-parametric function of $x_p$ in the form:
\begin{equation}
h(x_p,t) = \Phi(x_p(t))
\label{corfun}
\end{equation}
where $\Phi$ is a function of $x_p$ variable only.

In the following  we will explore the implications  that Eq (\ref{corfun})  may  have on partonic distributions of valence quarks.

\section{New Relations for Valence PDFs} 
 In general, PDFs depend on the two independent variables of $x$ and $t$ (or $Q^2$).  The relation of 
Eq.(\ref{corfun})  indicates that at the peak  the  $x$ weighted PDFs  depend analytically only on one variable, $x_p$,  and  
the $t$ dependence is expressed through  the  $x_p$'s dependence on $t$.  
This situation should result in specific relations for valence PDFs  at peak values. 

Starting with the relation:
\begin{equation}
h(x_p,t) = x_p\cdot q_V(x_p,t) = \Phi(x_p(t)),
\label{hPhi}
\end{equation}
we first consider the $t$ derivative using the right hand side of the above equation resulting in:
\begin{equation}
{dh(x_p,t)\over dt} = {d\Phi(x_p)\over d x_p}{d x_p\over d t}.
\label{dh_dt_Phi}
\end{equation}
Considering now the middle part of the Eq.(\ref{hPhi})  the $t$ derivative yields:
\begin{eqnarray}
& & {dh(x_p,t)\over dt}   = 
 {d (x_p(t) q_V(x_p,t))\over d t}  =  \nonumber  \\ 
 & & \ \ \   = {dx_p\over d t} q_V(x_p,t) + x_p {d q_V(x_p,t)\over d t} \nonumber  \\
& & = {dx_p\over d t} q_V(x_p,t) + x_p\left[{\partial q_V(x_p,t)\over \partial x_p} {d x_p\over d t} + {\partial q_V(x_p,t)\over \partial t}\right].
\nonumber \\
\label{dh_dt_xq}
\end{eqnarray}
Comparing Eqs.(\ref{dh_dt_Phi}) and (\ref{dh_dt_xq}) one obtains:
\begin{eqnarray}
x_p\left[{\partial q_V(x_p,t)\over \partial x_p}  {d x_p\over dt}\right.  &+& \left.{\partial q_V(x_p,t)\over \partial t}\right] = \nonumber \\
& & \left[ {d \Phi(x_p)\over d x_p} - q_V(x_p,t) \right] {d x_p\over dt}, \ \ \ \ 
\end{eqnarray}
or in a more compact form:
\begin{eqnarray}
\left[{\partial \log {q_V(x_p,t)}\over \partial x_p} + {1\over x_p}\right] + {\partial \log{q_V(x_p,t)}\over \partial t} /{d x_p\over dt} \nonumber \\
  = {d \log{\Phi(x_p)}\over d x_p}.
\label{newrel}
\end{eqnarray}
The above equation is rather unique since it  allows one to relate the $t$ derivative of the valence quark PDFs, which can be evaluated 
through the QCD evolution, to the $x_p$ derivative of the same distribution. 

Using various PDF sets
(e.g. \cite{Hou:2019efy,Dulat:2015mca,Accardi:2016qay}), which have been fitted to high energy data
such as deep inelastic scattering, Drell-Yan processes, etc., one can calculate numerically 
the LHS  of  Eq.(\ref{newrel}) and  thus evaluate the correlation 
function $\Phi(x_p)$  as a function of $t$ or $Q^2$.

\begin{figure}[th]
\centerline{ \includegraphics[width=8cm,height=6cm]{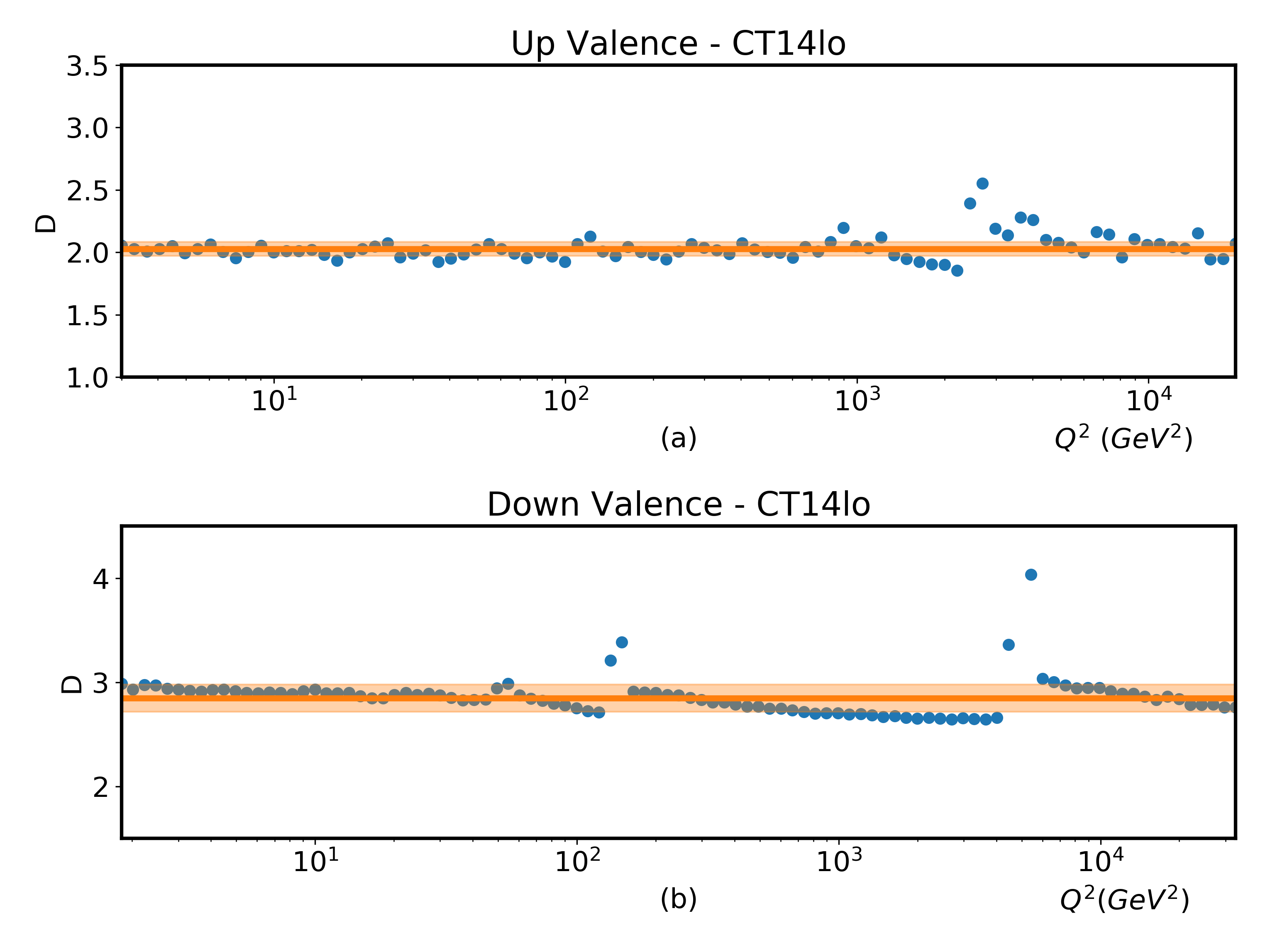}}
 \vspace{-0.2cm}
 \caption{(Color online.)   The data points are evaluations of LHS of Eq.(\ref{newrel}) using 
  CT14 lo \cite{Dulat:2015mca} for  valence quark distributions in the nucleon.  
 The factor D is the average value of evaluations with shaded area representing the standard deviation of calculated points. 
 The figure (a) is for valence $u$-quark and  (b)-for valence $d$- quark distributions.}
\label{dlogPhi_dx_lo}
\end{figure}

In Fig.\ref{dlogPhi_dx_lo} we present the calculation of the LHS of  Eq.(\ref{newrel}) 
using the CT14 PDF set published in Ref. \cite{Dulat:2015mca} at LO in $\alpha_s$ accessed via the LHAPDF library.  
Here we used  the QCD evolution equation to leading order to calculate ${\partial q_V(x_p,t)\over \partial t}$:
 \begin{eqnarray}
  & & \frac{\partial q_{V}(x,t)}{\partial t}  =    \frac{\alpha_s}{ 2\pi}\left\{ 2\left(1+\frac{4}{3}\log\left(1-x\right)\right) q_{V}(x,t) \right.
  \nonumber \\
  & & \left. +    \frac{4}{3}\int\limits_{x}^1\frac{dz}{1-z}\left(\frac{1+z^2}{ z}q_{V}\left(\frac{x}{z},t\right) - 2 q_{V}(x,t)\right)\right\},
\label{eveqlo}
\end{eqnarray}
while  ${d x_p \over d t}$ and ${\partial \log {q_V(x_p,t)}\over \partial x_p}$  are 
calculated numerically using valence d- and u-quark distributions at LO\cite{Dulat:2015mca}.

As Fig.\ref{dlogPhi_dx_lo} shows   the LHS  of Eq.(\ref{newrel}) is practically constant for 
 the all of $Q^2$s, covering four orders  of	 magnitude 
 ($1.8$~GeV$^2$ to  $3.3\times 10^4$~GeV$^2$). 
This  indicates that  ${d\log{\Phi(x_p)}\over dx_p} = D_{LO} = const$.

\begin{figure}[th]
\centerline{ \includegraphics[width=8cm,height=6cm]{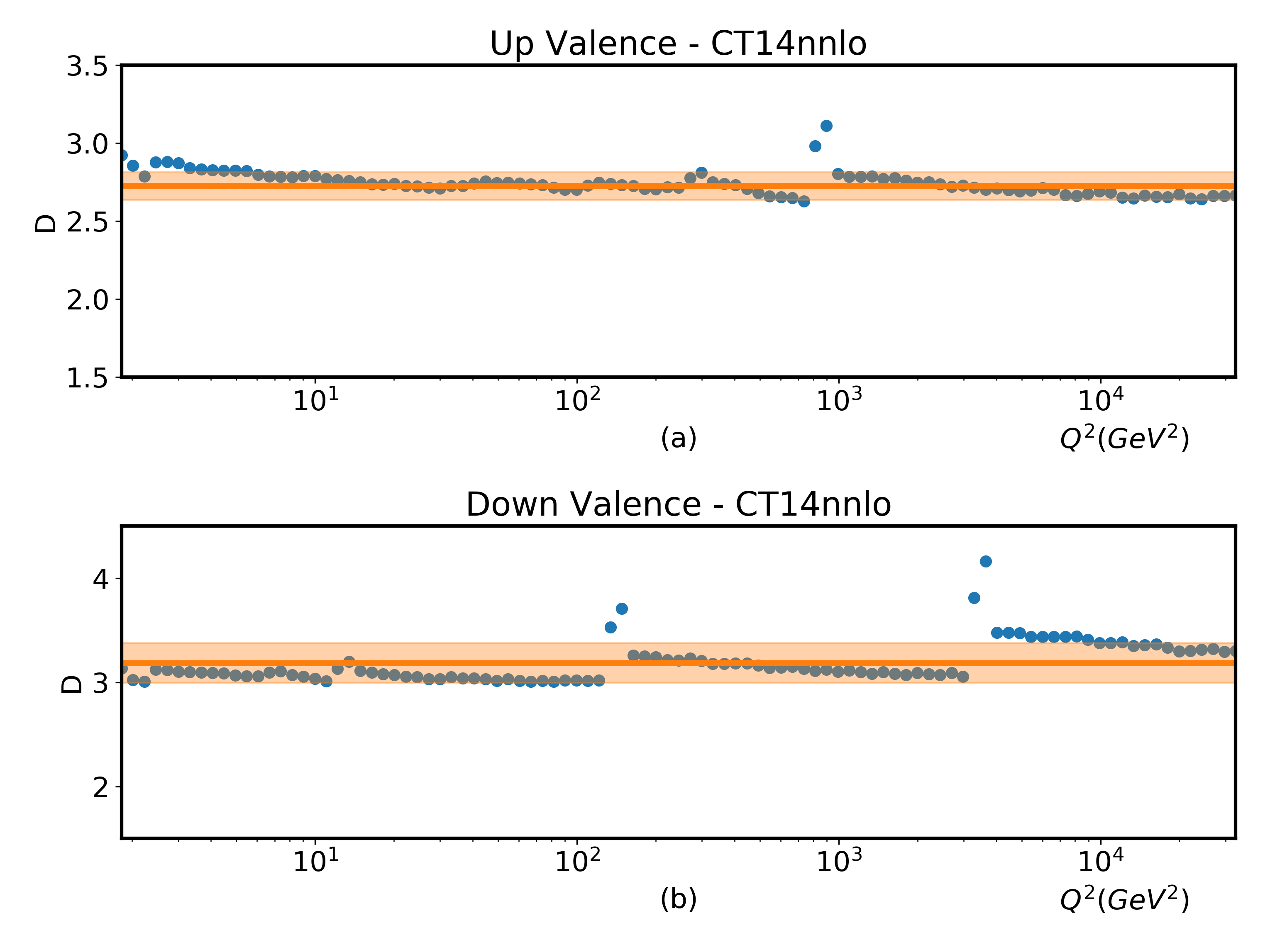}}
 \vspace{-0.2cm}
 \caption{(Color online.)   Same is in Fig.\ref{dlogPhi_dx_lo} but for next to next to leading order approximation for CT14 \cite{Dulat:2015mca}. The thick orange line is the $D$ obtained from averaging all the $D$'s for the central value curve, while the shaded orange region shows the region within $1 \sigma$ of the average.  Table~1 contains the PDF propagated errors.}
\label{dlogPhi_dx_nnl}
\end{figure}

In Fig.\ref{dlogPhi_dx_nnl} we present a similar evaluation for the  LHS of Eq.(\ref{newrel}) but for 
next-to-next-to leading order (NNLO).
As the figure shows, the LHS part  again produces an almost constant behavior for the entire range of $Q^2$.   
This  suggests again that   ${d\log{\Phi(x_p)}\over x_p}  = D_{NNL}  = const$. 
Overall Figs.(\ref{dlogPhi_dx_lo})  and (\ref{dlogPhi_dx_nnl})
show that the condition of   ${d\log{\Phi(x_p)}\over x_p} = const$ is approximately independent on the order of 
 approximation in the  QCD evolution.

The above observation of ${d\log{\Phi(x_p)}\over x_p}=const$ indicates that 
 the correlation function, $\Phi(x_p)$ has an exponential form:
\begin{equation}
h(x_p,t) = \Phi(x_p) = C e^{D x_p}
\label{expform}
\end{equation}
which is universal with respect to the order of  approximation in the QCD evolution equation, with differing exponents $D$ and overall factors $C$.
The values of $C$ and $D $ for $d$- and $u$- valence quarks in LO, NLO and NNLO approximations are  given in Table~I, which 
suggests that they converge with the increase of  approximation 
in the QCD evolution equation. Note that the LHAPDF gives PDF values by using spline interpolation between $(x,Q^2)$ grid points. This leads to instabilities when calculating PDF derivatives numerically (see \cite{Ball:2016spl}), hence the discontinuities and kinks in 
Fig. \ref{dlogPhi_dx_lo} and \ref{dlogPhi_dx_nnl}.

\begin{figure}[th]
 \vspace{-0.2cm}
\centerline{ \includegraphics[width=9cm,height=5cm]{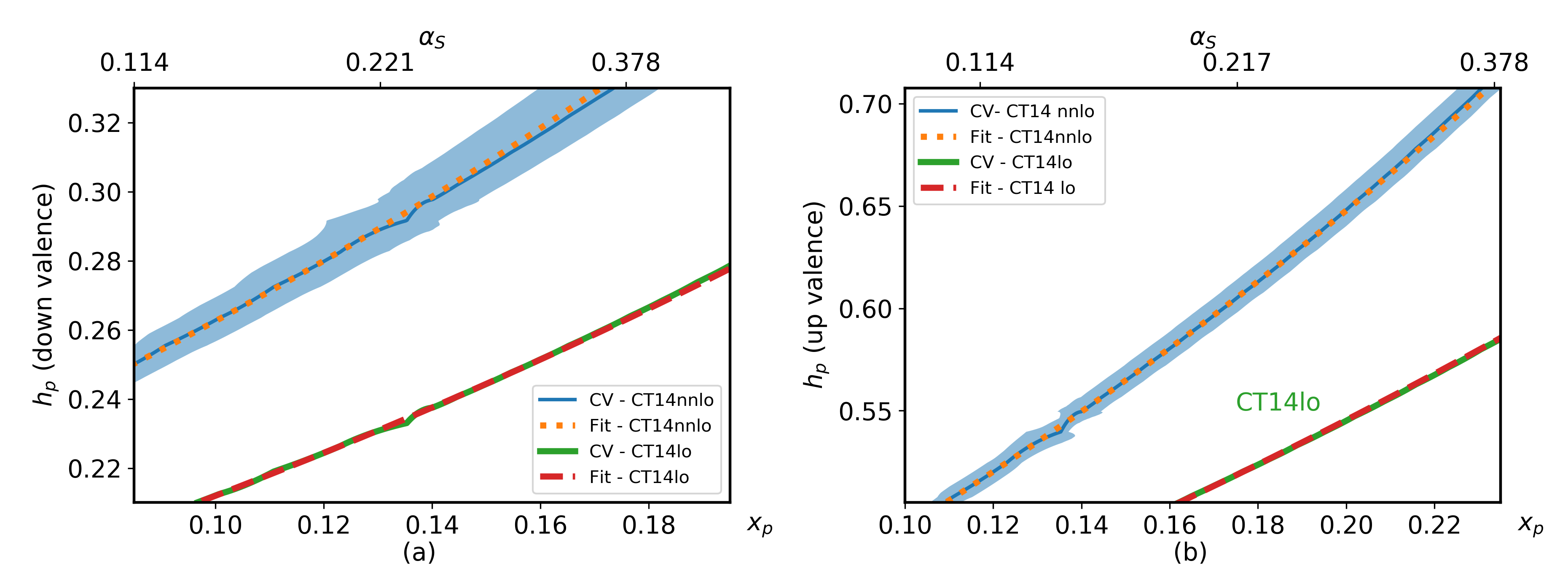}}
 \caption{(Color online.)   A comparison of the calculations (dotted lines) using  Eq.(\ref{expform}) with the actual 
 CT14 PDFs.
 Parementers $C$ and $D$ are given in Table I.  Solid lines represent the  central values and shaded area is the 
 Hessian error at 68\% confidence level of  the CT14nnlo parameterization. 
 Only the central value is shown for CT14lo.
 Figures (a) and (b) are for valence $d$- and $u$- quark distributions.
 The top $x$-axis shows the NNLO $\alpha_S$ evaluated at the same $Q^2$ as the corresponding $x_p$ on the bottom $x$-axis. }
\label{h_xp}
 \end{figure}

In Fig. \ref{h_xp} we use the parameters of Table I to compare the function of Eq.(\ref{expform}) with the $x_p$ dependence of 
$h(x_p,t)$ for 
down and up valence quarks in the nucleon in the LO and NNLO approximations for the CT14 parameterization.
The figure shows that indeed the $h(x_p,t)$ - $x_p$ correlation  follows almost ideally an  exponential form. 
It is worth mentioning that other modern PDF parameterizations 
using different ansatzs, renormalization and factorization schemes for PDFs give  similar results (see Section \ref{Other-PDF-and-Hadrons}).

\begin{table}
\label{CT14-values}
\caption{Parameters C and D of Eq.(\ref{expform}) for valence $d$- and 
$u$- quark distributions in the nucleon. Uncertainties are the 68\% confidence level and were obtained through 
(where appropriate) the PDF Hessian eigenvector set.}
\centering
\begin{tabular}{|c|c|c|c|}
\hline
Proton         & CT14lo              & CT14nlo            & CT14nnlo     \\
\hline
$d_V$ & 0.16,  2.8   & 0.19(1), 3.1(1)  & 0.193(6), 3.1(2)   \\
$u_V$ & 0.36,  2.02  & 0.37(1), 2.71(9) & 0.37(1), 2.7(1)  \\
\hline
\end{tabular}
\end{table}
\medskip
\medskip

Summarizing this chapter, we conclude that we found a new  empirical  relation for valence PDFs according to which the specific combination of 
the $t$ and $x_p$ derivatives of PDFs results into a constant value for all the range of $Q^2$ currently being discussed. The relation is:
\begin{eqnarray}
\left[{\partial \log {q_V(x_p,t)}\over \partial x_p} + {1\over x_p}\right] 
+ {\partial \log{q_V(x_p,t)}\over \partial t} {\big \slash}{d x_p\over dt} 
  = D
\label{newrello}
\vspace{-0.2cm}
\end{eqnarray}
where the constant $D$ depends on the flavor of the valence quark and order of  approximation in QCD evolution.

\section{Origin of the Exponential form of the Height-Position Correlation}
\label{OriginofExpForm}
Our observation of the exponential form of  Eq.~(\ref{expform})  in the current work is  purely empirical.
We used existing PDF   sets  in the given   approximation and estimated the expression in 
Eq.(\ref{newrel}) finding that it results in a constant number, $D$. It is an interesting problem to 
understand the origin of the observed  exponential form of  height-position correlation.
For this we note that 
while nonperturbative dynamics define the initial shape of the  valence PDFs, its change with $Q^2$ and therefore the 
height-position correlation is associated with QCD evolution and the baryonic sum rules that valence quarks must satisfy. Thus
 the change of the height of the peak of $h(x,t)$ function and its  position, $x_p$, is associated  with perturbative dynamics.

To  understand the origin of the exponential form of the ``height-position" correlation (Eq.(\ref{expform}))
one needs to consider the simultaneous solutions of evolution equations for $q_V(x,t)$  and $h^\prime(x_p,t)$  functions 
at the given  approximation  together with the condition of the maximum of the function 
${d h(x,t)\over dx}\mid_{x=x_p(t)} = 0$. The latter leads to 
the relation:
\begin{equation}
{dx_p \over dt} \left[ {\partial h^\prime (x_p,t)\over \partial x_p}\right] = -{\partial h^\prime(x_p,t)\over \partial t}
\label{fordxdt}
\end{equation}
where $h^\prime(x_p,t) = {d h(x,t)\over dx}\mid_{x=x_p(t)}$ and for the   ${\partial h^\prime(x_p,t)\over \partial t}$ term one can 
derive an evolution equation. For example, in LO approximation:
\begin{equation}
{\partial h^\prime(x_p,t)\over \partial t}  = {4\alpha_s\over 6\pi}\int\limits_{x_p}^{1} {dz\over 1-z} {1+z^2\over z} h^\prime ({x_p\over z},t).
\label{hprimeveq}
\end{equation}
The complexity of above equations (especially in NLO and NNLO approximations) makes it difficult to find an analytic solution in the form of Eq.(\ref{expform}).   
To understand the origin of such a correlation, 
currently the above equations are being studied  numerically\cite{Leon:inpro}  in 
different approximations,  using different ansatzs  for PDFs that describe experimental distributions at fixed $Q^2$.  These studies, which will be presented elsewhere,  indicate specific properties of evolution equation whose solutions furnish the correlation function in the form of 
Eq.~(\ref{expform}).

\section{The \boldmath{$x$} Dependence of \boldmath{$h(x,t)$} at the Vicinity of the Peak}

Using the fact that  $x_p< 1$,  from Eq.~(\ref{expform}) one observes 
that in the vicinity of $x\sim x_p$:
\begin{equation}
h(x,t) \equiv x q_V(x,t)  \approx  C + CDe x(1-x)^{1-x_p(t) \over x_p(t)}
\label{meanfield}
\end{equation}
where $C$ and $D$ are constants defined in Eq.(\ref{expform}) and $e$ is the Euler's number. 
It can be checked that the above function peaks at $x=x_p$ and 
its peak value corresponds to the  terms of the Taylor expansion of Eq.(\ref{expform}) in $x_p$ up to ${\cal O} (x_p^2)$.
It is interesting that even though the constant $D$ is due to dynamics of QCD evolution it defines also the valence PDF in 
the vicinity of the peak position, $x_p$.

As it follows from Eq.(\ref{meanfield}) the exponent of the $(1-x)$ term, ${1-x_p(t) \over x_p(t)}$ 
is defined by the position of the peak $x_p$. The latter, as we discussed above, changes continuously 
with $t$ due to QCD evolution.
The fact that the exponent of  $(1-x)$ term is not a constant and depends on the resolution of the probe  
indicates a more complex dynamics in the generation of  valence PDFs at $x\sim x_p$.  
For example, the decrease of $x_p$ with $t$ indicates 
that less momentum fraction is imparted to the interacting valence quarks, which can be due to the increase of the 
recoil mass of the nucleon, which itself is due to valence quark  radiation at large $t$.
Overall this observation indicates that no fixed number of constituent short-range interactions can be 
responsible for the dynamics of valence quark PDFs at any fixed values of $Q^2$ and at $x\sim x_p$.
 For a fixed number of exchanges,  one gets a constant exponent proportional to the number of exchanged particles\cite{Lepage:1980fj,Gunion:1983ay}.
On the other hand, a smoothly varying exponent can be  obtained by considering an effective interacting potential\cite{Leon:2020cev} 
in Weinberg type equations for relativistic bound states\cite{Weinberg:1966jm}.  Thus one may expect that  valence PDFs at $x\sim x_p$  at fixed $Q^2$ are generated by  mean-field type  interactions rather than by a combination of a finite number exchanges between valence quarks.

Note that only moving away from $x_p$  towards $x\to 1$ one 
should expect the mechanism of quark-quark  interaction through 
 a hard gluon exchange to become  important. Indeed the asymptotics of  empirical PDFs indicate that 
the exponent of $(1-x)$ part of the distribution approaches to a constant value at $x\ge 0.7-0.8$ (see e.g\cite{Dulat:2015mca,Ball:2016spl}).

In this respect the partonic dynamics for the 
valence sector can be similar to the one in the nuclear physics with mean field and short-range 
correlations dominating at different internal momentum regions of the constituents~ (see e.g. Ref.\cite{Sargsian:2012sm}).

\section{Other PDFs and Hadrons}  \label{Other-PDF-and-Hadrons}
\begin{figure}[th]
 \vspace{-0.1cm}
\centerline{ \includegraphics[width=9cm,height=6cm]{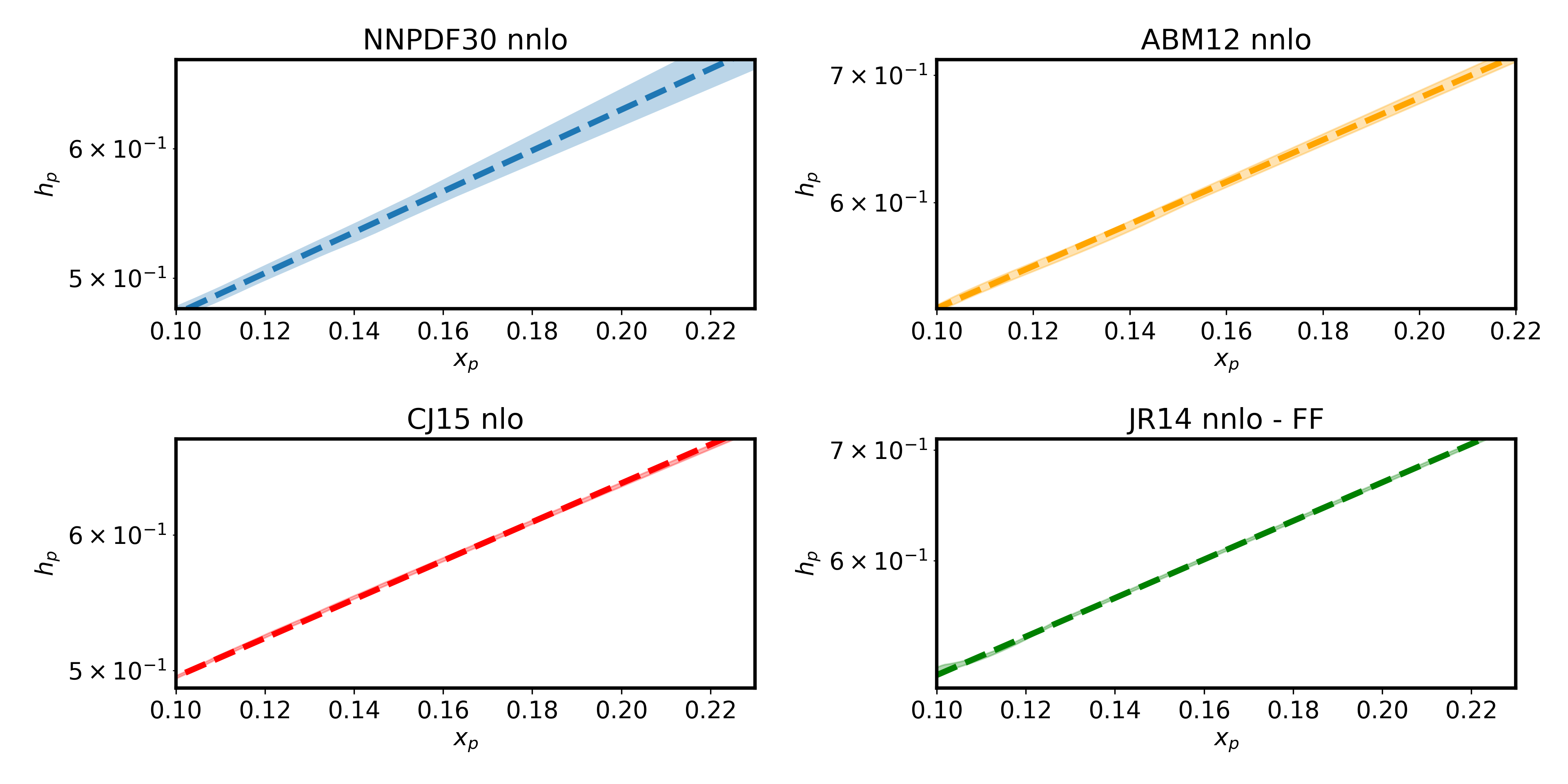}}
\vspace{-0.5cm}
 \caption{ (Color online.) The peak position-height relation for the PDF sets NNPDF 30 nnlo \cite{NNPDF:2014otw}, ABM12 nnlo \cite{Alekhin:2013nda},
CJ15 nlo \cite{Accardi:2016qay} and  JR14 nnlo - FF  \cite{Jimenez-Delgado:2014twa}  for the up valence. The shaded regions show the uncertainty at 68\% confidence level while the dashed curves are exponential fits for each PDF set.  }
 \label{pdf-several}
 \end{figure}

 \begin{figure}[th]
 \vspace{-0.4cm}
\centerline{ \includegraphics[width=9cm,height=6cm]{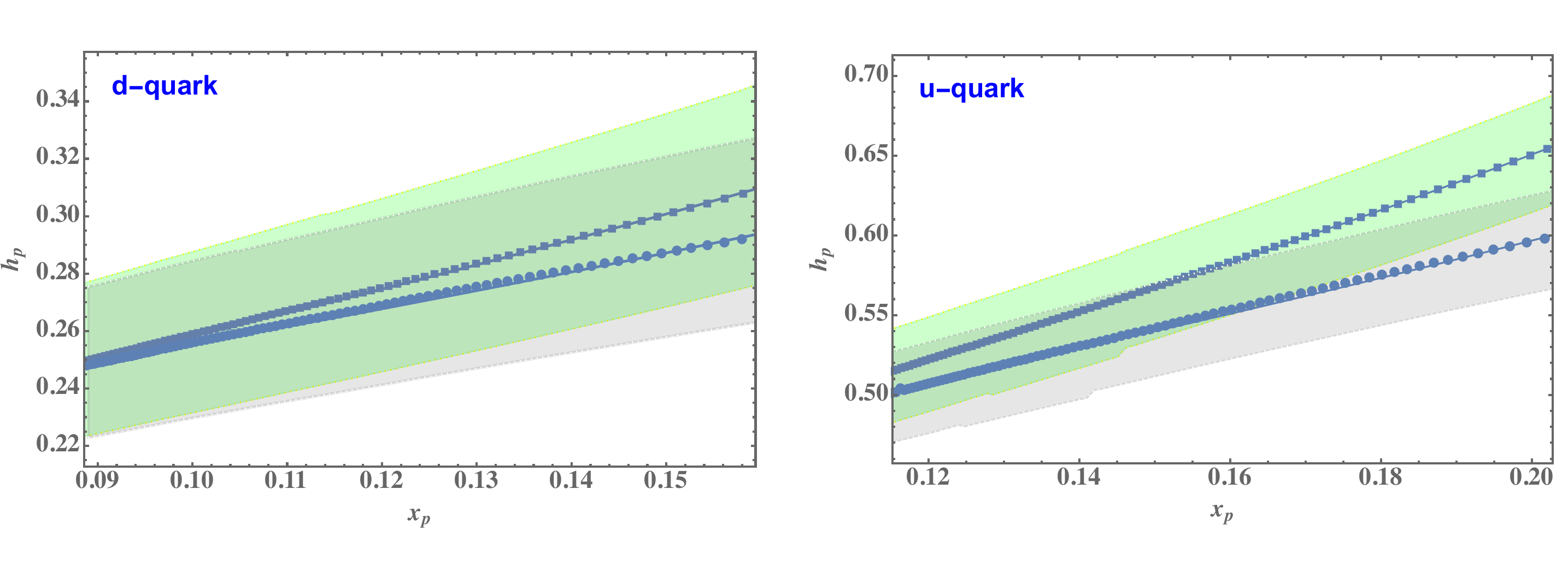}}
\vspace{-0.6cm}
 \caption{ (Color online.) The peak position-height relation for the 
 PDF sets  of CTEQ6 parameterization\cite{Pumplin:2002vw} for valence d- (left panel) and  valence u- (right panel)  
 quark qudistributions.  Solid squares indicate NLO parameterization based on the $\overline{MS}$ scheme and solid circles,  NLO based on DIS scheme\cite{Altarelli:1978id}. Curves are results of the fits according to Eq.(\ref{expform}).
The shaded regions show the  uncertainties evaluated by using error sets in Hessian representation presented in
Ref.\cite{Pumplin:2002vw}}.
 \label{MSbar_DIS}
 \end{figure}

The observed  ``height-position" correlation of the peak of the $h(x,t)$ function 
in the nucleon is a combination of two effects:  the dynamics that generate  the partonic distribution of valence quarks  at given $Q^2$
and the QCD evolution that shifts the strength of 
the distribution towards smaller $x$. 
The relation we found is  robust  for all   different PDF sets 
in leading order, which reinforces the expectation that effects is due to QCD evolution.  For the next to leading orders the 
PDF depends on factorization schemes, thus it is interesting to check whether the correlation of Eq.(\ref{expform}) persists 
in higher order approximations using different schemes.  Since all modern PDF parameterizations employ $\overline {MS}$ scheme\cite{ParticleDataGroup:2020ssz}  we first check  the validity of correlation for these sets  of PDFs. 
As can be seen in Fig. \ref{pdf-several} the PDF sets NNPDF 30 nnlo \cite{NNPDF:2014otw}, ABM12 nnlo \cite{Alekhin:2013nda},
CJ15 nlo \cite{Accardi:2016qay} and  JR14 nnlo - FF  \cite{Jimenez-Delgado:2014twa} all showed an exponential relation between the $h_p$  and $x_p$ for the up valence PDF.  In addition, the extracted $C$ and $D$ parameters were all similar to the one obtained from CT14 nlo and nnlo case. This is despite the fact that these sets use different orders of approximation and prescriptions. While the $\overline{MS}$ scheme is used in the DGLAP evolution equations the renormalization and factorization schemes used to deal with the heavy quark (which ends up impacting the light quark distributions via the momentum sum rule) differ. NNPDF uses the on-shell  renormalization scheme for the heavy quarks and FONLL-C for factorization, ABM12 nnlo and JR14 nnlo - FF both use $\overline{MS}$ for renormalization and FFNS ($n_f= 3$) for factorization, while CJ15 NLO uses the on-shell prescription and SACOT for factorization \cite{Accardi:2016ndt}.
They also differ in their initial ansatz, with NNPDF30 using neural networks and starting at $Q_0^2 = 1 \  \ GeV^2$, while ABM12 nnlo, CJ15 nlo, and JR14 nnlo - FF used simpler functional forms and started at $Q_0^2 = 0.5 ,1.69$ and $0.5 \ \ GeV^2$, respectively.

In Fig.\ref{MSbar_DIS} we compare available PDFs which are obtained using both $\overline{MS}$ and DIS\cite{Altarelli:1978id} schemes in next to leading order for CTEQ6 parameterization\cite{Pumplin:2002vw}.
Here we again observe the exponential form of correlation for both schemes, in which the results for DIS scheme is 
somewhat closer to that of  leading order approximation.  However, it is worth noticing  that total uncertainty for  CTEQ6 parametrization\cite{Pumplin:2002vw}  is larger than the uncertainties  for  modern parameterizations (e.g. Ref. \cite{Dulat:2015mca}).

From above comparisons  one might expect that correlation of type of Eq.(\ref{expform})  should be universal for 
valence PDFs  describing "experimental" distributions, satisfying  
specific sum rules, such as baryonic number sum rule for nucleons.   

\begin{figure}[t]
 \vspace{-0.3cm}
\centerline{ \includegraphics[width=8cm,height=5cm]{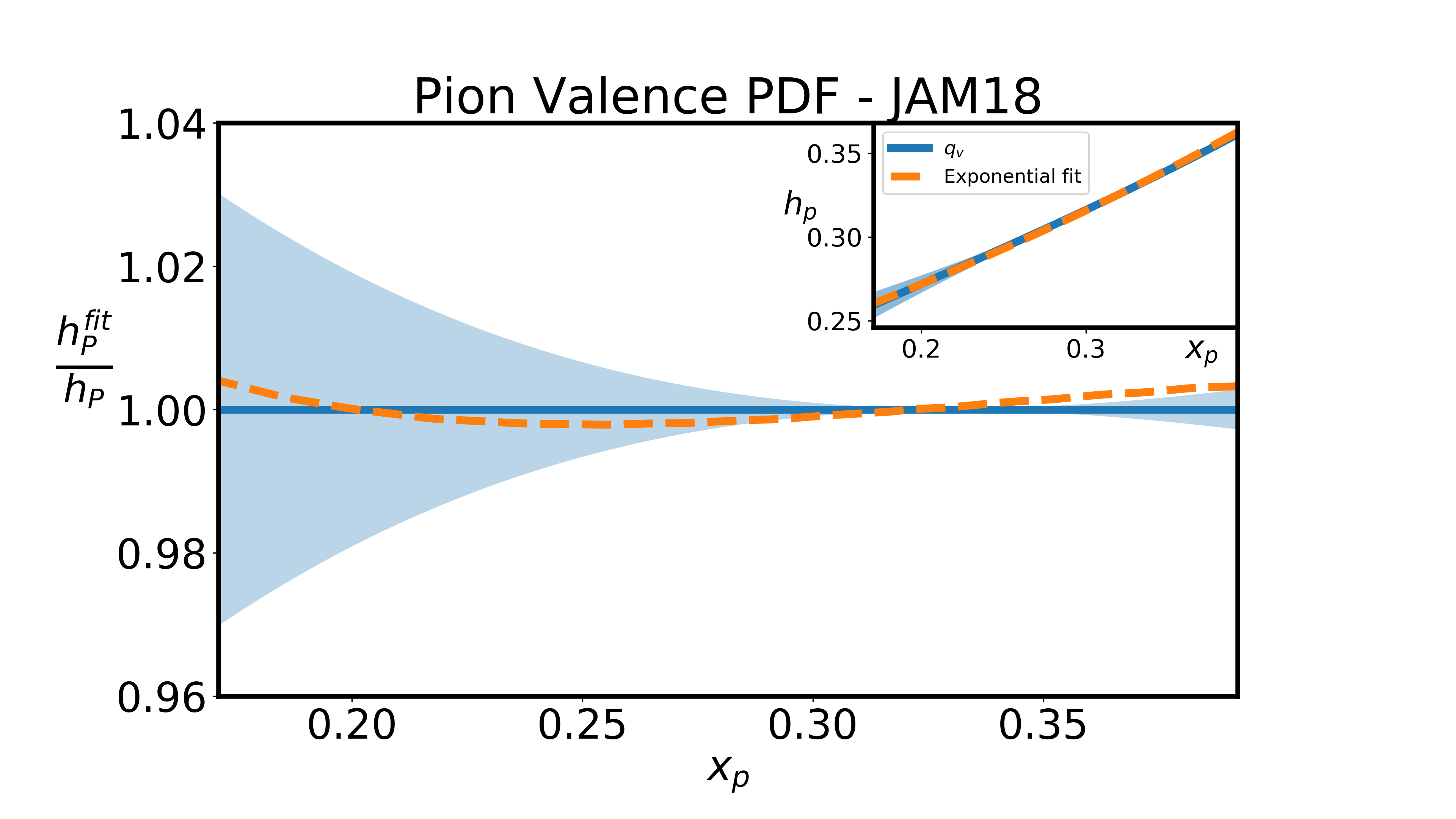}}
 \vspace{-0.5cm}
 \caption{(Color online.)  Ratio of exponential fit of the $x_p$ dependence of the $h(x_p)$ function using the parameterization of Ref.\cite{Barry:2018ort}.
 Dashed line is the fit according to Eq.(\ref{expform}), solid line central values and shaded area the error at 68 \%  confidence level of parameterization. The upper right hand corner shows the PDF $h_p$ function  vs $x_p$ as well as the fit. }
\label{h_xp_pion}
 \vspace{-0.3cm}
\end{figure}

It is interesting to explore the possibility of similar correlations  also for  mesons extending it to the sector of strange and charm quarks.
As an initial  result we also present  in Fig.\ref{h_xp_pion} the $x_p$ dependence of $x$ weighted valence quark distribution in the $\pi$-meson calculated based on 
the recently obtained PDF parameterization\cite{Barry:2018ort}. (Detailed analysis of pion PDFs will be presented elsewhere\cite{Leon:inpro}.)  
As the figure shows the exponential form of Eq.(\ref{expform}) fits the``height-position" correlation reasonably well with 
the parameters $C =0.202(3)$ and $D=1.50(2)$.  This represents a strong indication that the observed correlation (Eqs.(\ref{newrello}) 
and  (\ref{expform}))  is universal for any valence quark distribution in the hadron. 

\smallskip

\section{Conclusion and Outlook}
Concluding, we emphasize that the exponential form of the 
``height - position" correlation is a result of the specific relation 
between $x_p$ and $t$ derivatives of valence PDFs (Eq.(\ref{newrello})), which  results in a constant value $D$ for  all range of $Q^2$  that PDFs are considered.
The verification of this relation with PDF parameterizations obtained in LO, NLO and NNLO approximations  (e.g. Fig.\ref{dlogPhi_dx_lo} 
and Fig.\ref{dlogPhi_dx_nnl})  indicates that it is valid for any order of QCD coupling constant,
$\alpha_s$ based on $\overline{MS}$ scheme of factorization. Despite large uncertainties our analysis also indicates that 
Eq.(\ref{expform}) is valid  also for the other (DIS) scheme of factorization.
Such a universality of the exponential form of the correlation, in our view, is due to the dynamics of 
QCD evolution which can in principle  be studied as a separate analytical problem as discussed in Sec.\ref{OriginofExpForm}.

It will be interesting to verify the existence of relation (\ref{newrello}) for nuclear PDFs as well as for semi-inclusive  DIS processes  
sensitive to the valence quark distributions.
Relation (\ref{newrello}) can be used in the calculation of valence PDFs using lattice QCD, not just for the proton but for other hadrons whose PDFs are not well constrained by experiment.

Finally, one expects similar effects to be observed also for fragmentation functions, since evolution equations at least in LO have 
similar  splitting functions. In fact,  gluon fragmentation functions at small $x$ exhibit  features  reminiscent to the one discussed 
in this work (see e.g. Refs.\cite{Dokshitzer:1991wu,Ellis:1991qj} and references therein). 

Overall, establishing the universality of Eq.(\ref{newrello})  will allow to use it to constrain
$Q^2$ evolution of more complex  processes in higher order  approximations.

 \medskip
 \medskip

 {\bf Acknowledgments:}
This work is supported by U.S. DOE grant under contract DE-FG02-01ER41172.

\end{document}